\documentclass[sigconf, screen]{acmart}
\pdfoutput=1 
\AtBeginDocument{%
  \providecommand\BibTeX{{%
    \normalfont B\kern-0.5em{\scshape i\kern-0.25em b}\kern-0.8em\TeX}}}

\setcopyright{rightsretained}
\acmConference[COLIEE 2019]{COLIEE 2019 workshop: Competition on Legal
	Information Extraction/Entailment }{June 21, 2019}{Montreal, Quebec}
\acmISBN{}
\acmDOI{}

\usepackage{tikz}
\begin{document}

%
\title[Threshold-Based Retrieval and Textual Entailment Detection]{Threshold-Based Retrieval and Textual Entailment Detection on Legal Bar Exam Questions}

%
\author{Sabine Wehnert}
\email{sabine.wehnert@ovgu.de}
\affiliation{%
	\institution{Otto von Guericke University Magdeburg}
	\country{Germany}
}
\author{Sayed Anisul Hoque}
\email{sayed.hoque@st.ovgu.de}
\affiliation{%
	\institution{Otto von Guericke University Magdeburg}
	\country{Germany}
}
\author{Wolfram Fenske}
\email{wolfram.fenske@ovgu.de}
\affiliation{%
	\institution{Otto von Guericke University Magdeburg}
	\country{Germany}
}
\author{Gunter Saake}
\email{saake@iti.cs.uni-magdeburg.de}
\affiliation{%
  \institution{Otto von Guericke University  Magdeburg}
\country{Germany}
}

%
\renewcommand{\shortauthors}{Wehnert et al.}

\begin{abstract}
	Getting an overview over the legal domain has become challenging, especially in a broad, international context. Legal question answering systems have the potential to alleviate this task by automatically retrieving relevant legal texts for a specific statement and checking whether the meaning of the statement can be inferred from the found documents. We investigate a combination of the BM25 scoring method of Elasticsearch with word embeddings trained on English translations of the German and Japanese civil law. For this, we define criteria which select a dynamic number of relevant documents according to threshold scores. Exploiting two deep learning classifiers and their respective prediction bias with a threshold-based answer inclusion criterion has shown to be beneficial for the textual entailment task, when compared to the baseline. 
\end{abstract}

%
%


%

%
\maketitle

\section{Introduction}
Nowadays, globalization poses a challenge for many international organizations, since they need to ensure compliance to laws of all jurisdictions falling under the scope of their activities. Tracking changes in law is a challenging task, especially in statutory law where a single modification may affect the applicability of several legal articles, due to implicit co-dependencies between these documents. While domain experts are mostly required to ensure a reliable assessment of relationships among laws and their implications, the amount of legal documents is hard to oversee for a single person. Therefore, a decision support system can help in finding relevant laws and applying them to a specific question or statement.\footnote{The work is supported by Legal Horizon AG, Grant No.:1704/00082} Finding out whether a statement is true, given a corpus of legal text, falls under the task of legal question answering. A legal question answering system consists of two major parts: document retrieval and textual entailment recognition. In the retrieval phase, relevant law articles are selected for a query, having the form of a statement which shall be supported or contradicted by the law articles from the document collection. During the textual entailment phase, the query and accordingly retrieved legal documents are processed by a classification algorithm which returns ``yes'' in case of positive textual entailment or ``no'' otherwise. This work is a contribution to the Competition on Legal Information Extraction/Entailment (COLIEE) competition which provides a dataset from Japanese bar exam questions (translated to English) for evaluating the system performance on both tasks, retrieval and entailment classification.
Our contribution involves the following methods:
\begin{itemize}
\item We combine results from BM25 scoring with word embedding-based retrieval.
\item We develop a stacked encoder ensemble for entailment detection.
\item We use thresholding for both approaches. 
\end{itemize}

The remainder of this work is structured as follows: Section 2 outlines related work for both tasks with respect to their achievements using similar methods to our approach. In Section 3, we describe basic concepts for string representation in machine learning models, scoring methods and stacked encoders. We explain our approach in detail in Section 4 and show evaluation results in Section 5. After discussing those results, we conclude our findings and mention our future work considerations.

\section{Related Work}
The related work for our approach is divided in two parts: The legal information retrieval task and the entailment detection task. The first part consists of approaches using BM25 scoring or word embeddings, as well as similarity thresholding for a retrieval task. We further present deep learning methods, followed by approaches using thresholds for a textual entailment task.
\subsection{Legal Information Retrieval}
\subsubsection{BM25-Based Solutions}
In the COLIEE '16 competition, Onodera and Yoshioka apply BM25 scoring for information retrieval with several extensions using query keyword expansion. Their best result was an F-measure of 54.5\% \cite{kim2016coliee}. Arora et al. observe the best score with the BM25 scoring method on a different task of legal document retrieval \cite{arora2018}, compared to language models and term frequency - inverse document frequency (TF-IDF) weighting. This finding contradicts the previous observations from COLIEE competitions and the FIRE 2017 IRLeD Track, where ranking SVMs \cite{kim2015coliee} or language models \cite{mandal2017overview, tian2017hljit2017} performed better than mere BM25 scoring. Despite those observations, BM25 has shown to provide at least competitive results in many cases, so that we consider it as part of our approach. 
\subsubsection{Word Embeddings}
Word Embeddings have proven to be useful in many natural language processing contexts. We outline several works which have used this document feature representation for legal information retrieval. During the COLIEE '18 competition, the \texttt{SPABS} team was able to overcome vocabulary mismatch in some cases using an RNN-based solution with Word2Vec embeddings trained on English legal documents \cite{yoshiokaoverview}. Team UB used word embeddings with PL2 term weighting \cite{yoshiokaoverview}. Yoshioka et al. suggest to use semantic matching techniques for hard questions involving vocabulary mismatch combined with more reliable lexical methods for easy questions \cite{yoshiokaoverview}. This is the main motivation for our retrieval system, which incorporates lexical BM25 scoring and word embeddings as a semantic representation, respectively.
\subsubsection{Thresholding}
Thresholding based on similarity values can improve retrieval results by filtering out low-scoring matches. Islam and Inkpen use similarity thresholds to increase the precision of text matching \cite{islam2008semantic}. Stein et al. also employ thresholds for plagiarized document retrieval \cite{stein2007strategies}. In the COLIEE '18 competition, team \texttt{UBIRLED} use a similarity threshold for filtering out irrelevant case judgments \cite{kimcoliee}. Nanda et al. select the top-5 matching documents from a topic clustering approach \cite{nanda2017legal}. Given the document with the highest similarity score to the query, they apply thresholding, such that any further document will be incorporated into the result set if the distance to the topmost document is less than 15\%. Our approach uses a similar criterion for document inclusion.
\subsection{Legal Textual Entailment}
\subsubsection{Deep Learning Approaches}
Deep learning approaches have been used by several authors for entailment detection, starting with an application of a single-layered long short-term memory network (LSTM) for input encoding by Bowman et al. \cite{bowman2015large}. The encoded features from both texts are concatenated and passed through three 200-dimensional tanh layers to a softmax classifier for predicting the entailment relationship. The task is performed on the SNLI\footnote{\texttt{nlp.stanford.edu/projects/snli/}} dataset which is based on image captioning. Rockt\"{a}schel et al. apply neural attention \cite{Bahdanau2015NeuralMT} for entailment recognition on the same SNLI corpus \cite{Rocktschel2016ReasoningAE}. Two LSTM networks are employed for encoding the query and the document, whereby the output vectors from the document are used by an attention mechanism for each word in the respective query. Their method achieves 83.5\% accuracy, which is compared to the results by Bowman et al. an improvement of 3.3 percentage points. Liu et al. use a bidirectional LSTM with an attention mechanism \cite{Liu2016LearningNL} and obtained 85\% accuracy on the SNLI dataset. A stacked encoder architecture developed by Nie and Bansal achieved 86.1\% accuracy on the SNLI dataset. Considering that result as the state-of-the-art, we adapt the main idea to our task in the legal domain and further explain this architecture in section \ref{subsec:SE}. Do et al. use a convolutional neural network (CNN) with word
embeddings \cite{DoNTNN17}. They incorporate additional features from a TF-IDF and latent
semantic indexing (LSI) representation of the sentences. Finally, they feed these features in conjunction with the output of the CNN model into a multi-layer perceptron (MLP) network to predict the answer. 

We are inspired by the work of the Chen et al., which focuses on a factoid question and answering system \cite{chen2017reading}. Their goal is to predict a sequence in the document to answer the query, as opposed to our task of detecting an entailment relationship. They trained two multi-layer bi-directional LSTMs to encode the articles and the query. For encoding the article, they extract multiple features from the query and document pairs: word embeddings of the document (300-dimensional Glove embeddings), an exact matching flag, token features (part-of-speech tags, named entity tags, normalized term frequencies) and attention scores for the similarity of a document and the aligned query. These features are concatenated to form the input vector for the LSTM that encodes the article. The question is encoded without extracting any features. Their evaluation is based on the top five pages returned by the algorithm, and results in {77.8\%} of correct answers on the SQuAD \cite{rajpurkar2016squad} dataset.

Nanda et al. apply a hybrid network of LSTM networks coupled with
a CNN, with the final prediction based on a softmax classifier \cite{nanda2017legal}. They use pre-trained general-purpose word embeddings from the Google news corpus, consisting of 3 billion words. Their accuracy for the COLIEE '17 competition was 53.8\%, which they attribute to the general-purpose embeddings which may not capture important semantic relationships needed for the legal domain.

From these works, we conclude that LSTM architectures are suitable for entailment detection for open-domain tasks. However, the COLIEE dataset poses a challenge for deep learning models due to the specific meaning of terms in the legal domain and the rather small size of the dataset. Therefore, we refrain from training word embeddings on the statute law competition corpus only, but consider using other general-purpose word embeddings and a slightly different architecture compared to the previous work. We also find in the related work that extracting additional features from the documents can improve the classifier performance.

\subsubsection{Thresholding}
Thresholding for the entailment task is applied in two cases: First, the entailment detection can be done by using a similarity threshold. This works similar to an attention layer in a neural network, which applies a focus on a subsection of the input. The second case refers to thresholding in classifier output probabilities.
 During the COLIEE '15 competition, Kano obtained good results using a threshold for snippet scoring of the entailment task in the runs \texttt{Kanolab1}, \texttt{Kanolab2} and \texttt{Kanolab3} \cite{kim2015coliee}. Ha et al. show that similarity scoring can be used for entailment detection \cite{ha2012refining}, however, this may only work for low-level inferences. Instead, Carvalho et al. suggest to use classifiers with a rich feature set \cite{carvalho2015lexical}. We incorporate an exact matching component for obvious cases of positive entailment in our solution and train a classifier for the majority of cases. Glickman et al. tune their probabilistic entailment detection method with a threshold \cite{glickman2005web}. Rooney et al. note that for entailment detection, exploiting individual classifier bias with a voting scheme or stacking is common \cite{rooney2014investigation}. We employ a voting scheme which is based on using thresholds on the output probabilities of classifiers in our ensemble and consider their respective bias.
\section{Background}
In this section, we introduce the required concepts to understand the components of our approach. First, we describe two text representation approaches. Second, we present scoring methods for the retrieval task. Finally, we explain a method for sequence encoding and decoding which is frequently applied in open-domain question answering. 
\subsection{Text Representation} 
In order to apply machine learning techniques on text data, a suitable representation of the input is needed. We consider two alternative approaches: the bag-of-words model and word embeddings. The bag-of-words representation collects all words of a document regardless of their order as assigns each distinct word to a unique index. For two texts to be considered as similar, there needs to be a significant lexical overlap. 
In contrast, word embeddings are semantic representations of a word, which do not require a lexical overlap. However, these representations need to be created, for instance from a neural network which is trained on a reference corpus to predict the next probable word in a sequence. A model for training word embeddings is Word2Vec \cite{mikolov2013distributed}. The Word2Vec algorithm learns the word embeddings
by using the continuous bag-of-words (CBOW) model or the continuous skip-gram
model. The embeddings are learned in the CBOW model by predicting the current
word based on a context window. In contrast, the continuous skip-gram model predicts the surrounding context given a reference word. General-purpose pre-trained word embeddings are often used, for instance Glove embeddings \cite{pennington2014glove}. During the embedding training, the hidden layers of this network capture information such as the co-occurrence of two words which are present in the same context. The resulting n-dimensional weights from the hidden layer of the network are then called word embeddings. It is important in this context to choose a good reference corpus which represents the vocabulary and common word use in the domain of the task well to obtain adequate embedding weights. 
\subsection{Weighting and Scoring Methods} 
Regardless of the chosen representation method, there can be the need to adjust the weight of each word while aggregating to the document level. A common approach for weighting is a discounting of the term frequency within a document by its overall frequency within the corpus. This term frequency - inverse document frequency (TF-IDF) weighting scheme takes into account that only certain words describe a document well. That is, words which occur frequently within a document, but relatively seldom across the rest of the document collection are considered to be keywords and obtain the highest weight. Likewise, words which occur often in the whole corpus are treated as stopwords and their weight is diminished by the denominator term of the inverse document frequency.
Generally, retrieval systems often contain a ranking function, where the similarity or a relevance score between the query and candidate documents is computed in order to determine document relevance. We describe two scoring methods in more detail: BM25 similarity scoring and Word Centroid Distance, whereby the latter treats the analogous problem of distance measurement for similarity scoring. 
BM25 scoring, also referred to as Okapi BM25, is similar to TF-IDF weighting, however it limits the influence of the term frequency and common words between query and document, following a notion of term eliteness as a poisson distribution \cite{robertson2009probabilistic}.
The Word Centroid Distance (WCD) has been specifically developed for word embeddings by Kusner et al. \cite{kusner2015word}. It depicts the minimum cumulative distance that the words of a document need to travel to reach all words of another document in the embedding space. Thereby, a semantic distance metric is provided, given that word embeddings exist which capture those semantic relationships between the words of the respective domain. As a result, paraphrasing documents which convey the same meaning without sharing a single word can be still detected as a match.
\subsection{Sequence Encoding and Decoding}\label{subsec:SE}
Using sentences or article paragraphs as an input for a machine learning algorithm requires a transformation to a lower-dimensional feature representation. Since the order of words and thereby also dependencies among words can impact the entailment relationship, we briefly introduce an approach to model text sequences: the sequence-to-sequence (Seq2Seq) model by Sutskever et al. \cite{sutskever2014sequence}. It consists of a recurrent neural network (RNN) as an encoder, which maps the input to a fixed-length context vector. This context vector is then accessed by the decoder RNN to generate the target sequence. This approach is popular in the general question answering domain \cite{reddy2018coqa}. The Seq2Seq model has been extended by Bahdanau et al. by an attention layer, so that a smaller segment of the fixed-length vector can be used by the decoder to predict the target \cite{Bahdanau2015NeuralMT}. Neural network architectures built for question answering encode the query and the document separately. A common practice is to concatenate both context vectors to a single context representation and separate them by markers of question start and sequence end for the final prediction \cite{strohquestion}. 
The question answering problem we consider in this work is reduced to a binary output of confirming or rejecting the statement from the query. Therefore, the decoder is replaced by an entailment classifier, as in the work by Nie and Bansal \cite{nie2017shortcut}. While they use an MLP layer followed by a softmax layer for the prediction, we employ a multi-layer neural network, with stepwise-linear activation functions, the rectified linear units (ReLUs), due to their accuracy and simplicity at the same time. In particular, recent studies have investigated the approximation capability of ReLUs in hidden layers, and found that there is a gain in accuracy, although the nonlinearity of weights is given up \cite{schmidt2017nonparametric, yarotsky2018optimal}. Schmidt-Hieber names desirable properties of the ReLU function: First, the projection property of ReLUs can be exploited to propagate a signal over layers without distortion to synchronize smaller subnetworks of varying depth, and second, the upper bound of all network parameters equals one, thus limiting the amplitude of weight updates during training \cite{schmidt2017nonparametric}. 
To conclude, sequence encoding has been applied on several question answering problems and therefore, we consider it for the task of legal bar exam entailment classification. In the next section, we present the details of our approach.

\section{Threshold-Based Retrieval and Entailment Detection}
In this section, we present our methodology to retrieve relevant laws for a given query. Then, we show our approach for classifying the entailment relationship.
\subsection{Law Retrieval combining BM25 Scoring and Word Embeddings}
Previous competition results suggest that it can be beneficial to combine the advantage of lexical matching (using the bag-of-words approach) with more recent semantic matching techniques (e.g., word embeddings) \cite{kimcoliee}. 
\subsubsection{Conceptual Approach}
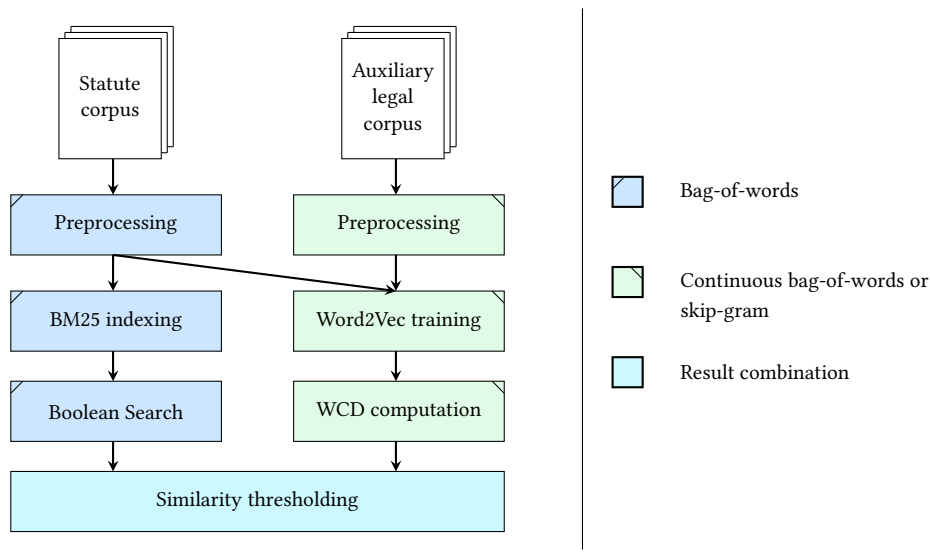
\begin{figure*}[ht]

	\centering
	\colorlet{blue}{blue!50!cyan!50}
\colorlet{turquoise}{blue!40!cyan!50}
\colorlet{light}{blue!40!white}
\colorlet{bluegreen}{blue!30!green!70}
\colorlet{lightgreen}{bluegreen!30}
\colorlet{petrol}{teal!60!cyan!40}
\colorlet{light2}{petrol!30!white}
\usetikzlibrary{patterns}
\usetikzlibrary{shapes.multipart}

\begin{tikzpicture}[scale=0.8, every node/.style={ align=center, scale=0.9}]

\begin{scope}[]
\draw (-7.7,6) rectangle (-6,4)node[pos=.5]{Statute \\ corpus};
\node [outer sep=0,inner sep=0,minimum size=0](v1) at (-6.8,4) {};
\node[outer sep=0,inner sep=0,minimum size=0] (v10) at (-7.6,6.1) {};
\node [outer sep=0,inner sep=0,minimum size=0](v11) at (-5.9,6.1) {};
\node [outer sep=0,inner sep=0,minimum size=0](v12) at (-5.9,4.1) {};
\node[outer sep=0,inner sep=0,minimum size=0] (v13) at (-7.6,6) {};
\node [outer sep=0,inner sep=0,minimum size=0](v14) at (-6,4.1) {};
\node [outer sep=0,inner sep=0,minimum size=0](v15) at (-5.8,6.2) {};
\node [outer sep=0,inner sep=0,minimum size=0](v16) at (-5.8,4.2) {};
\node[outer sep=0,inner sep=0,minimum size=0] (v17) at (-7.5,6.1) {};
\node [outer sep=0,inner sep=0,minimum size=0](v18) at (-5.9,4.2) {};
\node[outer sep=0,inner sep=0,minimum size=0] (v19) at (-7.5,6.2) {};
\node [outer sep=0,inner sep=0,minimum size=0](v20) at (-5.8,6.2) {};

\end{scope}

\begin{scope}[]
\draw (-3,6) rectangle (-1.3,4)node[pos=.5]{Auxiliary \\ legal  \\ corpus};
\node [outer sep=0,inner sep=0,minimum size=0](v21) at (-2.8,4) {};
\node[outer sep=0,inner sep=0,minimum size=0] (v210) at (-2.9,6.1) {};
\node [outer sep=0,inner sep=0,minimum size=0](v211) at (-1.2,6.1) {};
\node [outer sep=0,inner sep=0,minimum size=0](v212) at (-1.2,4.1) {};
\node[outer sep=0,inner sep=0,minimum size=0] (v213) at (-2.9,6) {};
\node [outer sep=0,inner sep=0,minimum size=0](v214) at (-1.3,4.1) {};
\node [outer sep=0,inner sep=0,minimum size=0](v215) at (-1.1,6.2) {};
\node [outer sep=0,inner sep=0,minimum size=0](v216) at (-1.1,4.2) {};
\node[outer sep=0,inner sep=0,minimum size=0] (v217) at (-2.8,6.1) {};
\node [outer sep=0,inner sep=0,minimum size=0](v218) at (-1.2,4.2) {};
\node[outer sep=0,inner sep=0,minimum size=0] (v219) at (-2.8,6.2) {};
\node [outer sep=0,inner sep=0,minimum size=0](v220) at (-1.1,6.2) {};

\end{scope}

\draw [fill=light] (-8.5,1.8) rectangle (-5,0.8)node[pos=.5]{BM25 indexing};
\draw [fill=light] (-8.5,0.3) rectangle (-5,-0.7)node[pos=.5]{Boolean Search};
\draw [fill=light] (-8.5,3.4) rectangle (-5,2.4)node[pos=.5]{Preprocessing};
\draw [fill=bluegreen!20] (-3.8,3.4) rectangle (-0.3,2.4)node[pos=.5]{Preprocessing};
\draw [fill=bluegreen!20] (-3.8,1.8) rectangle (-0.3,0.8)node[pos=.5]{Word2Vec training};
\draw [fill=bluegreen!20] (-3.8,0.3) rectangle (-0.3,-0.7)node[pos=.5]{WCD computation};
\draw [fill=turquoise!50] (-8.5,-1.2) rectangle (-0.3,-2.2)node[pos=.5]{Similarity thresholding};

\node [outer sep=0,inner sep=0,minimum size=0] (v2) at (-6.8,3.4) {};
\node [outer sep=0,inner sep=0,minimum size=0] (v3) at (-2.1,1.8) {};
\draw  [-stealth, thick](v1) edge (v2);

\node [outer sep=0,inner sep=0,minimum size=0] (v34) at (-2.1,0.8) {};
\node [outer sep=0,inner sep=0,minimum size=0] (v35) at (-2.1,0.3) {};
\draw  [-stealth, thick](v34) edge (v35);

\node [outer sep=0,inner sep=0,minimum size=0] (v4) at (-6.8,2.4) {};
\node [outer sep=0,inner sep=0,minimum size=0] (v5) at (-6.8,1.8) {};
\draw   [-stealth, thick](v4) edge (v5);
\draw  [-stealth, thick](v4) edge (v3);
\node [outer sep=0,inner sep=0,minimum size=0] (v8) at (-6.8,0.8) {};
\node [outer sep=0,inner sep=0,minimum size=0] (v9) at (-6.8,0.3) {};
\draw   [-stealth, thick](v8) edge (v9);
\node [outer sep=0,inner sep=0,minimum size=0](v6) at (1,6.5) {};
\node [outer sep=0,inner sep=0,minimum size=0](v7) at (1,-2.5) {};
\draw  (v6) edge (v7);
\draw [thick, fill = light] (1.5,3.7) rectangle (2,3.2);
\draw [thick, fill=bluegreen!20] (1.5,2.2) rectangle (2,1.7);
\draw [thick, fill=turquoise!50] (1.5,0.7) rectangle (2,0.2);
\node [right] at (2.5,3.45) {Bag-of-words};
\node [right] at (2.5,1.95) {Continuous bag-of-words or};
\node [right] at (2.5,1.45) {skip-gram};
\node [right] at (2.5,0.45) {Result combination};

\draw  (v10) edge (v11);
\draw  (v11) edge (v12);

\draw  (v13) edge (v10);

\draw  (v14) edge (v12);

\draw  (v15) edge (v16);

\draw  (v16) edge (v18);

\draw  (v19) edge (v17);
\draw  (v19) edge (v20);

\draw  (v210) edge (v211);
\draw  (v211) edge (v212);

\draw  (v213) edge (v210);

\draw  (v214) edge (v212);

\draw  (v215) edge (v216);

\draw  (v216) edge (v218);

\draw  (v219) edge (v217);
\draw  (v219) edge (v220);

\node [outer sep=0,inner sep=0,minimum size=0] (v22) at (-2.1,2.4) {};
\draw  [-stealth, thick](v22) edge (v3);
\node [outer sep=0,inner sep=0,minimum size=0] (v32) at (-2.1,4) {};
\node [outer sep=0,inner sep=0,minimum size=0] (v33) at (-2.1,3.4) {};
\draw  [-stealth, thick](v32) edge (v33);
\node [outer sep=0,inner sep=0,minimum size=0] (v37) at (-2.1,-0.7) {};
\node [outer sep=0,inner sep=0,minimum size=0] (v38) at (-2.1,-1.2) {};
\draw  [-stealth, thick](v37) edge (v38);
\node [outer sep=0,inner sep=0,minimum size=0] (v39) at (-6.8,-0.7) {};
\node [outer sep=0,inner sep=0,minimum size=0] (v40) at (-6.8,-1.2) {};
\draw  [-stealth, thick](v39) edge (v40);
\node [outer sep=0,inner sep=0,minimum size=0] (v29) at (-8.5,3.2) {};
\node [outer sep=0,inner sep=0,minimum size=0] (v30) at (-8.3,3.4) {};
\node  [outer sep=0,inner sep=0,minimum size=0](v31) at (-8.5,1.6) {};
\node [outer sep=0,inner sep=0,minimum size=0] (v36) at (-8.3,1.8) {};
\node [outer sep=0,inner sep=0,minimum size=0] (v42) at (-8.3,0.3) {};
\node  [outer sep=0,inner sep=0,minimum size=0](v41) at (-8.5,0.1) {};
\node [outer sep=0,inner sep=0,minimum size=0] (v23) at (-0.5,3.4) {};
\node  [outer sep=0,inner sep=0,minimum size=0](v24) at (-0.3,3.2) {};
\node  [outer sep=0,inner sep=0,minimum size=0](v25) at (-0.5,1.8) {};
\node [outer sep=0,inner sep=0,minimum size=0] (v26) at (-0.3,1.6) {};
\node  [outer sep=0,inner sep=0,minimum size=0](v27) at (-0.5,0.3) {};
\node [outer sep=0,inner sep=0,minimum size=0] (v28) at (-0.3,0.1) {};
\draw  (v23) edge (v24);
\draw  (v25) edge (v26);
\draw  (v27) edge (v28);
\draw  (v29) edge (v30);
\draw  (v31) edge (v36);
\draw  (v41) edge (v42);
\node [outer sep=0,inner sep=0,minimum size=0] (v44) at (1.7,3.7) {};
\node [outer sep=0,inner sep=0,minimum size=0] (v43) at (1.5,3.5) {};
\node  [outer sep=0,inner sep=0,minimum size=0](v45) at (1.8,2.2) {};
\node [outer sep=0,inner sep=0,minimum size=0] (v46) at (2,2) {};
\draw  (v43) edge (v44);
\draw  (v45) edge (v46);
\end{tikzpicture}
	\caption{Overview of the information retrieval workflow.}
	\Description{Information retrieval workflow overview.}
	\label{fig:ir}
\end{figure*}
Our retrieval system uses lexical BM25 scoring together with Word2Vec embeddings. An overview of the process is shown in Figure \ref{fig:ir}. During the preprocessing phase, we separate the articles in the Japanese civil law file by using regular expressions. Then, we create an index over all articles and apply the BM25 scoring method. In addition, we implement a separate pipeline for word embedding creation from the German civil code as an auxiliary corpus. Preliminary experiments have shown that solely using Word2Vec on the English translation of the Japanese civil code does not provide a good embedding quality, so we enriched the corpus for the embedding creation by an English translation of the German civil code\footnote{\texttt{https://www.gesetze-im-internet.de/englisch\_bgb/}}. With our trained word embeddings, we applied the Word Centroid Distance (WCD) as a scoring function. Further experiments led us to apply TF-IDF weighting on the Word Centroid Distance to increase the number of relevant documents. We find that both methods returned similar results, however, the word embeddings can contribute correct documents in case of vocabulary mismatch and sufficient semantic similarity. Therefore, we introduced similarity thresholding and prioritize the word embedding-based matching documents over the ones which scored high only on BM25. Since the number of relevant documents varies by query, we define criteria for similarity thresholds in both methods to select one or multiple documents, if there is a high similarity to the query. Setting a high similarity threshold results in a high precision, but a low recall. We therefore select all documents with a high similarity automatically as relevant. Additional documents are only included if they fulfill the threshold criteria which we set manually to optimize retrieval performance on the training data. For instance, a top-1 document with high confidence has either a BM25 score that is at least twice as high as the following second document and larger than the length of the query with an added constant of 20, or that document has a word embedding-based similarity of 90\%. The length of the query is considered as a criterion for the BM25 results because the lexical overlap should be significant and not fall below 20\% similarity score. Further criteria for lower confidence results are based on the same logic, just with relaxed parameters. Depending on the number of already accepted predictions from higher confidence votes, we define criteria for further result inclusion. For example, if there are no predictions from high, medium or low similarity thresholds, the top-1 prediction without any confidence is retrieved from the results of the word embedding-based similarity scoring.

\subsubsection{Implementation Details}
Our implementation for the retrieval task is based on Python 3 and Elasticsearch 6.4.1. Furthermore we use the library \texttt{vec4ir}\footnote{\texttt{https://github.com/lgalke/vec4ir/}} provided by Galke et al. for word embedding-based retrieval and re-weighting frequent terms with the inverse document frequency before the centroids are calculated for the WCD \cite{galke2017word}. The 300-dimensional Word2Vec embeddings are trained with the continuous bag-of-words model and a context window size of 5. We train the embeddings for 700 epochs on the English version of the German Civil Code and then for further 800 epochs on the English version of the Japanese Civil Code. The input string is tokenized using the spaCy\footnote{\texttt{https://spacy.io/}} library.
\subsection{Textual Entailment using an ensemble of Stacked LSTM Encoders}
\subsubsection{Conceptual Approach}
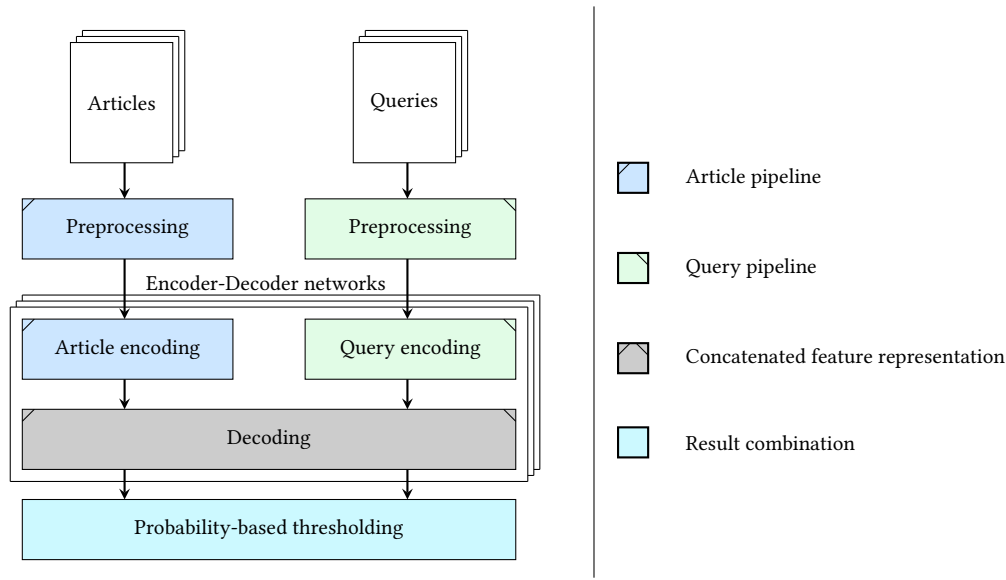
\begin{figure*}[ht]
   
   \centering
  \colorlet{blue}{blue!50!cyan!50}
\colorlet{turquoise}{blue!40!cyan!50}
\colorlet{light}{blue!40!white}
\colorlet{bluegreen}{blue!30!green!70}
\colorlet{lightgreen}{bluegreen!30}
\colorlet{petrol}{teal!60!cyan!40}
\colorlet{light2}{petrol!30!white}
\usetikzlibrary{patterns}
\usetikzlibrary{shapes.multipart}
\begin{tikzpicture}[scale=0.8, every node/.style={ align=center, scale=0.9}]

\begin{scope}[shift={(0,0.4)}]
\draw (-7.7,6) rectangle (-6,4)node [pos=.5] {Articles};
\node [outer sep=0,inner sep=0,minimum size=0] (v1) at (-6.8,4) {};
\node[outer sep=0,inner sep=0,minimum size=0] (v10) at (-7.6,6.1) {};
\node [outer sep=0,inner sep=0,minimum size=0] (v11) at (-5.9,6.1) {};
\node [outer sep=0,inner sep=0,minimum size=0] (v12) at (-5.9,4.1) {};
\node[outer sep=0,inner sep=0,minimum size=0] (v13) at (-7.6,6) {};
\node [outer sep=0,inner sep=0,minimum size=0] (v14) at (-6,4.1) {};
\node [outer sep=0,inner sep=0,minimum size=0] (v15) at (-5.8,6.2) {};
\node [outer sep=0,inner sep=0,minimum size=0] (v16) at (-5.8,4.2) {};
\node[outer sep=0,inner sep=0,minimum size=0] (v17) at (-7.5,6.1) {};
\node [outer sep=0,inner sep=0,minimum size=0] (v18) at (-5.9,4.2) {};
\node[outer sep=0,inner sep=0,minimum size=0] (v19) at (-7.5,6.2) {};
\node [outer sep=0,inner sep=0,minimum size=0] (v20) at (-5.8,6.2) {};

\end{scope}

\begin{scope}[shift={(0,0.4)}]
\draw (-3,6) rectangle (-1.3,4)node [pos=.5] {Queries};
\node [outer sep=0,inner sep=0,minimum size=0] (v21) at (-2.8,4) {};
\node[outer sep=0,inner sep=0,minimum size=0] (v210) at (-2.9,6.1) {};
\node [outer sep=0,inner sep=0,minimum size=0] (v211) at (-1.2,6.1) {};
\node [outer sep=0,inner sep=0,minimum size=0] (v212) at (-1.2,4.1) {};
\node[outer sep=0,inner sep=0,minimum size=0] (v213) at (-2.9,6) {};
\node [outer sep=0,inner sep=0,minimum size=0] (v214) at (-1.3,4.1) {};
\node [outer sep=0,inner sep=0,minimum size=0] (v215) at (-1.1,6.2) {};
\node [outer sep=0,inner sep=0,minimum size=0] (v216) at (-1.1,4.2) {};
\node[outer sep=0,inner sep=0,minimum size=0] (v217) at (-2.8,6.1) {};
\node [outer sep=0,inner sep=0,minimum size=0] (v218) at (-1.2,4.2) {};
\node[outer sep=0,inner sep=0,minimum size=0] (v219) at (-2.8,6.2) {};
\node [outer sep=0,inner sep=0,minimum size=0] (v220) at (-1.1,6.2) {};

\end{scope}

\draw [fill=light] (-8.5,1.8) rectangle (-5,0.8)node[pos=.5]{Article encoding};
\draw [fill=black!20] (-8.5,0.3) rectangle (-0.3,-0.7)node[pos=.5]{Decoding};
\draw [fill=light] (-8.5,3.8) rectangle (-5,2.8)node[pos=.5]{Preprocessing};
\draw [fill=bluegreen!20] (-3.8,3.8) rectangle (-0.3,2.8)node[pos=.5]{Preprocessing};
\draw [fill=bluegreen!20] (-3.8,1.8) rectangle (-0.3,0.8)node[pos=.5]{Query encoding};

\draw [fill=turquoise!50] (-8.5,-1.2) rectangle (-0.3,-2.2)node[pos=.5]{Probability-based thresholding};

\node [outer sep=0,inner sep=0,minimum size=0] (v2) at (-6.8,3.8) {};
\node [outer sep=0,inner sep=0,minimum size=0] (v3) at (-2.1,1.8) {};
\draw  [-stealth, thick](v1) edge (v2);

\node [outer sep=0,inner sep=0,minimum size=0] (v34) at (-2.1,0.8) {};
\node [outer sep=0,inner sep=0,minimum size=0] (v35) at (-2.1,0.3) {};
\draw  [-stealth, thick](v34) edge (v35);

\node [outer sep=0,inner sep=0,minimum size=0] (v4) at (-6.8,2.8) {};
\node [outer sep=0,inner sep=0,minimum size=0] (v5) at (-6.8,1.8) {};
\draw   [-stealth, thick](v4) edge (v5);

\node [outer sep=0,inner sep=0,minimum size=0] (v8) at (-6.8,0.8) {};
\node [outer sep=0,inner sep=0,minimum size=0] (v9) at (-6.8,0.3) {};
\draw   [-stealth, thick](v8) edge (v9);
\node [outer sep=0,inner sep=0,minimum size=0](v6) at (1,7) {};
\node [outer sep=0,inner sep=0,minimum size=0](v7) at (1,-2.5) {};
\draw  (v6) edge (v7);
\draw [thick, fill = light] (1.4,4.4) rectangle (1.9,3.9);
\draw [thick, fill=bluegreen!20] (1.4,2.9) rectangle (1.9,2.4);
\draw [thick, fill=black!20] (1.4,1.4) rectangle (1.9,0.9);
\node [right] at (2.4,4.15) {Article pipeline};
\node [right] at (2.4,2.65) {Query pipeline};

\node [right] at (2.4,1.15) {Concatenated feature representation};

\draw [thick, fill=turquoise!50] (1.4,-0.5) rectangle (1.9,0);
\node [right] at (2.4,-0.25) {Result combination};

\draw  (v10) edge (v11);
\draw  (v11) edge (v12);

\draw  (v13) edge (v10);

\draw  (v14) edge (v12);

\draw  (v15) edge (v16);

\draw  (v16) edge (v18);

\draw  (v19) edge (v17);
\draw  (v19) edge (v20);

\draw  (v210) edge (v211);
\draw  (v211) edge (v212);

\draw  (v213) edge (v210);

\draw  (v214) edge (v212);

\draw  (v215) edge (v216);

\draw  (v216) edge (v218);

\draw  (v219) edge (v217);
\draw  (v219) edge (v220);

\node [outer sep=0,inner sep=0,minimum size=0] (v22) at (-2.1,2.8) {};
\draw  [-stealth, thick](v22) edge (v3);
\node [outer sep=0,inner sep=0,minimum size=0] (v32) at (-2.1,4.4) {};
\node [outer sep=0,inner sep=0,minimum size=0] (v33) at (-2.1,3.8) {};
\draw  [-stealth, thick](v32) edge (v33);
\node [outer sep=0,inner sep=0,minimum size=0] (v37) at (-2.1,-0.7) {};
\node [outer sep=0,inner sep=0,minimum size=0] (v38) at (-2.1,-1.2) {};
\draw  [-stealth, thick](v37) edge (v38);
\node [outer sep=0,inner sep=0,minimum size=0] (v39) at (-6.8,-0.7) {};
\node [outer sep=0,inner sep=0,minimum size=0] (v40) at (-6.8,-1.2) {};
\draw  [-stealth, thick](v39) edge (v40);
\node [outer sep=0,inner sep=0,minimum size=0] (v23) at (-8.5,3.6) {};
\node [outer sep=0,inner sep=0,minimum size=0] (v24) at (-8.3,3.8) {};
\draw  (v23) edge (v24);
\node [outer sep=0,inner sep=0,minimum size=0](v25) at (-8.5,1.6) {};
\node[outer sep=0,inner sep=0,minimum size=0] (v26) at (-8.3,1.8) {};
\draw  (v25) edge (v26);
\node [outer sep=0,inner sep=0,minimum size=0](v27) at (1.4,4.2) {};
\node[outer sep=0,inner sep=0,minimum size=0] (v28) at (1.6,4.4) {};
\draw  (v27) edge (v28);
\node [outer sep=0,inner sep=0,minimum size=0] (v29) at (-0.5,3.8) {};
\node[outer sep=0,inner sep=0,minimum size=0] (v30) at (-0.3,3.6) {};
\draw  (v29) edge (v30);
\node [outer sep=0,inner sep=0,minimum size=0](v31) at (-0.5,1.8) {};
\node [outer sep=0,inner sep=0,minimum size=0](v36) at (-0.3,1.6) {};
\draw  (v31) edge (v36);
\node [outer sep=0,inner sep=0,minimum size=0](v41) at (-0.5,0.3) {};
\node[outer sep=0,inner sep=0,minimum size=0] (v42) at (-0.3,0.1) {};
\node [outer sep=0,inner sep=0,minimum size=0](v43) at (-8.3,0.3) {};
\node [outer sep=0,inner sep=0,minimum size=0](v44) at (-8.5,0.1) {};
\draw  (v41) edge (v42);
\draw  (v43) edge (v44);
\node [outer sep=0,inner sep=0,minimum size=0](v45) at (1.7,2.9) {};
\node[outer sep=0,inner sep=0,minimum size=0] (v46) at (1.9,2.7) {};
\node [outer sep=0,inner sep=0,minimum size=0](v49) at (1.6,1.4) {};
\node[outer sep=0,inner sep=0,minimum size=0] (v50) at (1.4,1.2) {};
\node[outer sep=0,inner sep=0,minimum size=0] (v47) at (1.7,1.4) {};
\node [outer sep=0,inner sep=0,minimum size=0](v48) at (1.9,1.2) {};
\draw  (v45) edge (v46);
\draw  (v47) edge (v48);
\draw  (v49) edge (v50);
\node[outer sep=0,inner sep=0,minimum size=0] (v51) at (-8.7,2) {};
\node[outer sep=0,inner sep=0,minimum size=0](v54) at (-0.1,2) {};
\node [outer sep=0,inner sep=0,minimum size=0] (v52) at (-8.7,-0.9) {};
\node[outer sep=0,inner sep=0,minimum size=0] (v53) at (-0.1,-0.9) {};
\draw  (v51) edge (v52);
\draw  (v52) edge (v53);
\draw  (v53) edge (v54);
\draw  (v54) edge (v51);
\node [outer sep=0,inner sep=0,minimum size=0](v56) at (-8.6,2.1) {};
\node[outer sep=0,inner sep=0,minimum size=0] (v57) at (0,2.1) {};
\node[outer sep=0,inner sep=0,minimum size=0] (v58) at (0,-0.8) {};
\node[outer sep=0,inner sep=0,minimum size=0] (v59) at (-0.1,-0.8) {};
\node [outer sep=0,inner sep=0,minimum size=0](v55) at (-8.6,2) {};
\draw  (v55) edge (v56);
\draw  (v56) edge (v57);
\draw  (v57) edge (v58);
\draw  (v58) edge (v59);
\node at (-4.45,2.4) {Encoder-Decoder networks};
\node[outer sep=0,inner sep=0,minimum size=0] (v61) at (-8.5,2.2) {};
\node [outer sep=0,inner sep=0,minimum size=0](v60) at (-8.5,2.1) {};
\node [outer sep=0,inner sep=0,minimum size=0](v62) at (0.1,2.2) {};
\node [outer sep=0,inner sep=0,minimum size=0](v63) at (0.1,-0.7) {};
\node[outer sep=0,inner sep=0,minimum size=0] (v64) at (0,-0.7) {};
\draw  (v60) edge (v61);
\draw  (v61) edge (v62);
\draw  (v62) edge (v63);
\draw  (v63) edge (v64);
\end{tikzpicture}
  \caption{Overview of the textual entailment workflow.}
  \Description{Textual entailment workflow overview.}
  \label{fig:ent}
\end{figure*}

Our approach for entailment detection consists of stacked encoders and an entailment classifier, inspired by the stacked bidirectional LSTM encoder by Nie and Bansal \cite{nie2017shortcut}. Since there may be identical sequences of query and document, such easy cases are returned as positive entailment beforehand. The remaining query-document pairs are processed by the stacked encoder approach. An overview of our classification process is shown in Figure \ref{fig:ent}.
As a first step, the queries and articles are preprocessed by tokenizing the sequences. Both have a separate pipeline, since four additional features are extracted from the article. The first three features are obtained by comparing the article with query tokens and storing matches as a binary feature if a match occurs in the original form, in the lower-cased form or in the lemmatized token form, respectively. We also compute the normalized term frequency for each term in the article. The main advantage of using the normalized term frequency is to lessen the effect of
high occurrences of a term in a text sequence \cite{manning2010introduction}. Once extracted, these four features along with the tokenized form of the article and the
query become the input to the deep learning model. The feature vectors of query and article are passed to the stacked encoders separately. The first layer in the pipeline of the deep learning model is the embedding layer. Due to a few spelling mistakes, some words in the documents could not be mapped to the Glove embeddings. Because of the small corpus size, we defined a dictionary with correct term spellings. If the Glove embedding is not found, the correct surrogate word is looked up, at least for cases we detected in the training dataset. We then incorporate an attention layer because of the different size of the query and article length (with the article being potentially longer). Another reason for choosing an attention layer is the interpretability of the learned input characteristics. Words causing higher neuron activation can be traced from the attention scores of the model. As a next layer, this network uses a deep bidirectional long short-term memory (LSTM) network and outputs a fixed-length context representation vector. The context vectors from both the article and the query encoder are concatenated and processed by the entailment classifier consisting of stacked linear layers with ReLU activation functions. The binary output is generated by a sigmoid function providing the predicted entailment relationship. Afterwards, models with the highest accuracy are investigated with respect to their bias to set rules for the ensemble voting mechanism. We use the probability of each class - positive or negative entailment - as an indicator of classifier confidence. For example, if one model predicts positive entailment while the other model votes for negative entailment, the positive result is chosen if the probability of the positive class in the approving model is larger than 53\%.

\subsubsection{Implementation Details}
The entailment recognition task is also implemented using Python 3. For preprocessing we again use spaCy. The deep learning models are built by using PyTorch\footnote{\texttt{https://pytorch.org}}. We train two models using different random seed values for 100 epochs with a batch size of 20 and an Adamax optimizer with the cross entropy loss. There are 3 bidirectional LSTM layers in both encoders - for query and article features - with a hidden size of 100. We apply a dropout of 20\% for each bidirectional LSTM layer and use gradient clipping. To generate the prediction, there are 4 stacked linear layers with ReLU activation functions, followed by a sigmoid function for the final output.

\section{Results}
The evaluation results are based on the assessment of the COLIEE competition. First, we present and discuss the results of the retrieval task. Second, we proceed with the results of the textual entailment task and insights from the assessment. While a complete question answering system is evaluated based on the whole pipeline - retrieval and entailment classification on the retrieved documents - we decided to solve both tasks independently from each other by taking the gold standard set of all relevant documents for the entailment recognition. 
\subsection{Law Retrieval}
We summarize our document coverage for high, medium and low confidence in the training dataset in Table \ref{tab:cov}. The BM25 scoring method achieves a higher coverage than the word embedding-based solution and has therefore a slightly worse retrieval performance on the high, medium and low confidence thresholds. We decide to prioritize word embedding-based results over the BM25 scoring when the similarity - thus also the confidence - is high.
\begin{table}[ht]
	
	\caption{Retrieval coverage and performance by using similarity thresholds for several confidence levels. The number of documents retrieved by using BM25-based similarity thresholding is denoted as N$_{\text{bm25}}$, the respective word embedding similarity-based document count as N$_{\text{emb}}$. The metrics coverage \emph{C}, precision \emph{P}, recall \emph{R} and F2-measure \emph{F2} are depicted in percentages.}
	\label{tab:cov}
	\begin{tabular}{lrrr}
			\toprule
		Metric&High&Medium&Low\\
		\midrule
		N$_{\text{bm25}}$&89&390&453\\
		N$_{\text{emb}}$&53&154&380\\
		C$_{\text{bm25}}$&8.1&35.8&41.7\\
		C$_{\text{emb}}$&4.9&14.0&31.6\\
		P$_{\text{bm25}}$&97.7&87.5&77.6\\
		P$_{\text{emb}}$&100.0&94.9&81.0 \\
		R$_{\text{bm25}}$&94.5&79.5&66.3\\
		R$_{\text{emb}}$&94.3&86.9&78.0\\
	    F2$_{\text{bm25}}$&94.2&79.5&66.9\\
	    F2$_{\text{emb}}$&94.8&87.2&76.1\\
		\bottomrule
	\end{tabular}
\end{table}
For the combined result on the training dataset (H18-H29), we achieve a precision of 29.3\%, a recall of 63.59\% and an F2-measure of 48.2\%. Preliminary experiments have shown that our approach outperforms simple TF-IDF scoring in Elasticsearch on the training data, as well. 
Table \ref{tab:ir} contains our result in the COLIEE competition compared to the competitors with the highest and the lowest score.
\begin{table}[ht]
	\caption{Retrieval results of our team DBSE for the metrics F2-measure \emph{F2}, precision \emph{P}, recall \emph{R}, mean average precision \emph{MAP}, recall at 5 \emph{R@5},  recall at 10 \emph{R@10} and recall at 30 \emph{R@30}, calculated as macro-averaged values and depicted in percentages.}
	\label{tab:ir}
	\begin{tabular}{lrrr}
		\toprule
		Metric&UA-TFIDF&\textbf{DBSE}&iitptfidf\\
		\midrule
		F2&54.93&\textbf{46.59}&40.08\\
		P&59.18&\textbf{45.44}&43.88\\
		R&54.42&\textbf{49.32}&39.63\\
		MAP&61.81&\textbf{51.19}&50.56\\
		R@5&61.98&\textbf{51.24}&57.02\\
		R@10&69.42&\textbf{61.98}&62.81\\
		R@30&76.03&\textbf{66.94}&75.21\\
		\bottomrule
	\end{tabular}
\end{table}

Similarly, on the macro-average evaluation setup, we achieve an F2-score of 46.59\%, while other participants scored from 40.08\% to 54.93\%. However, results from the competition runs indicate that our Word2Vec embeddings cannot guarantee relevant document detection, since our recall at 30 has been outperformed by all other methods. The Word2Vec embeddings may be trained on a larger corpus of legal text - not only civil codes - to provide more valuable semantic representations. Preliminary experiments of using the skip-gram model instead of CBOW for training word embeddings decreased all scores on the training dataset, so that we only selected embeddings based on the CBOW model for result submission. 

\subsection{Textual Entailment}
We selected two models with the highest accuracy. They are both biased towards the negative class, although we divided the dataset in such a way that there are more positive examples in the training dataset (with a validation set from H28). We obtain from both models 63.6\% accuracy on the validation set. We consider the models to be complementary because they predict positive entailment on different instances. Therefore, we combine their predictions. 
\begin{table}[ht]
	\caption{Entailment classification results of our team DBSE for the accuracy \emph{A} metric, calculated as macro-averaged value and depicted in percentages.}
	\label{tab:ent}
	\begin{tabular}{lrrr}
		\toprule
		Metric&UA\_Ex&\textbf{DBSE}&Baseline\\
		\midrule
		A&68.37&\textbf{57.14}&52.04\\
		\bottomrule
	\end{tabular}
\end{table}

\begin{figure*}[ht]
	\centering
	\includegraphics[width=0.9\textwidth]{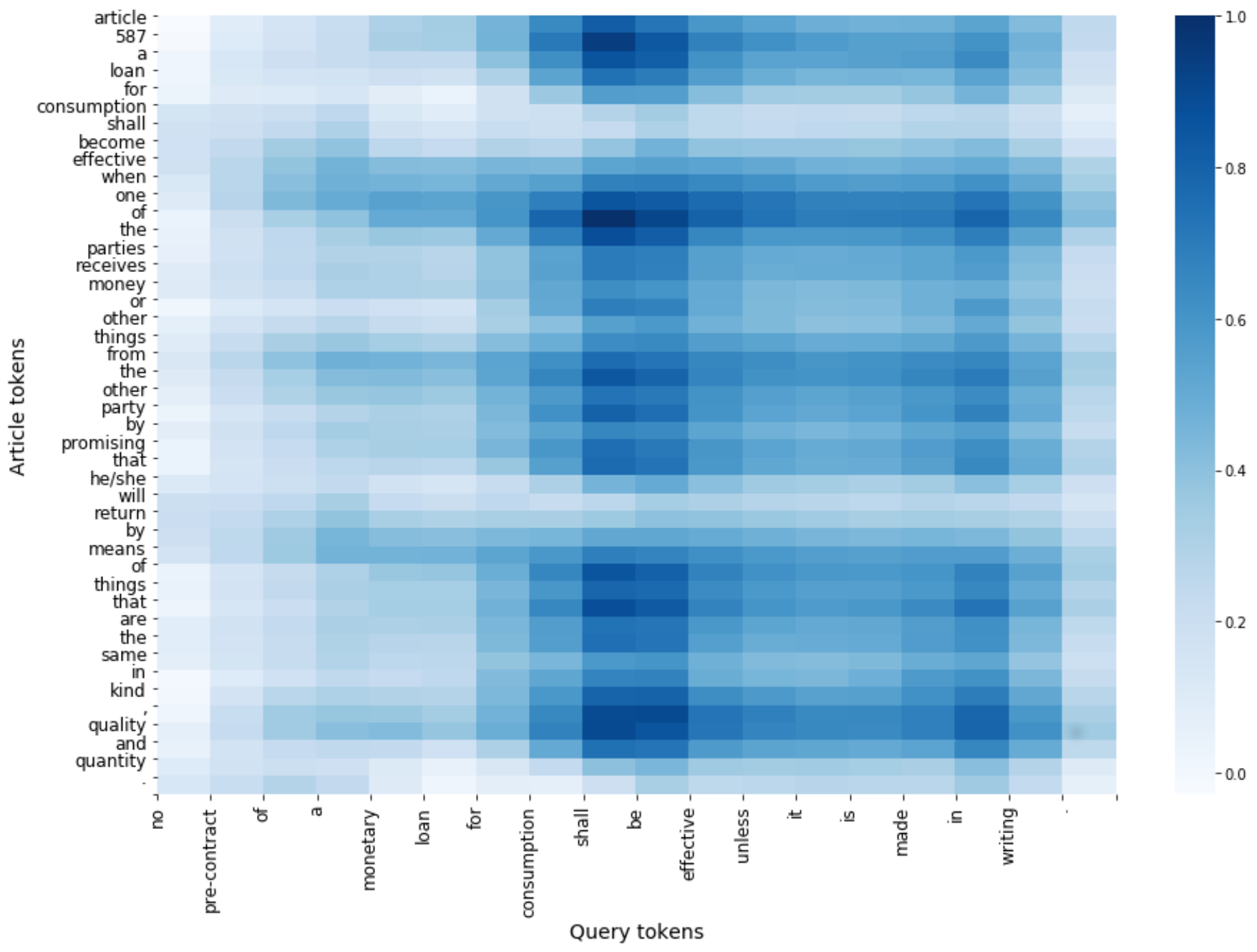}
	\caption{Neural attention example for task H30-23-A. Our model incorrectly predicted a negative entailment relationship to be positive.}
	\Description{Neural attention example for task H30-23-A.}
	\label{fig:att23a}
\end{figure*}
\begin{figure*}[ht]
	\centering
	\includegraphics[width=0.9\textwidth]{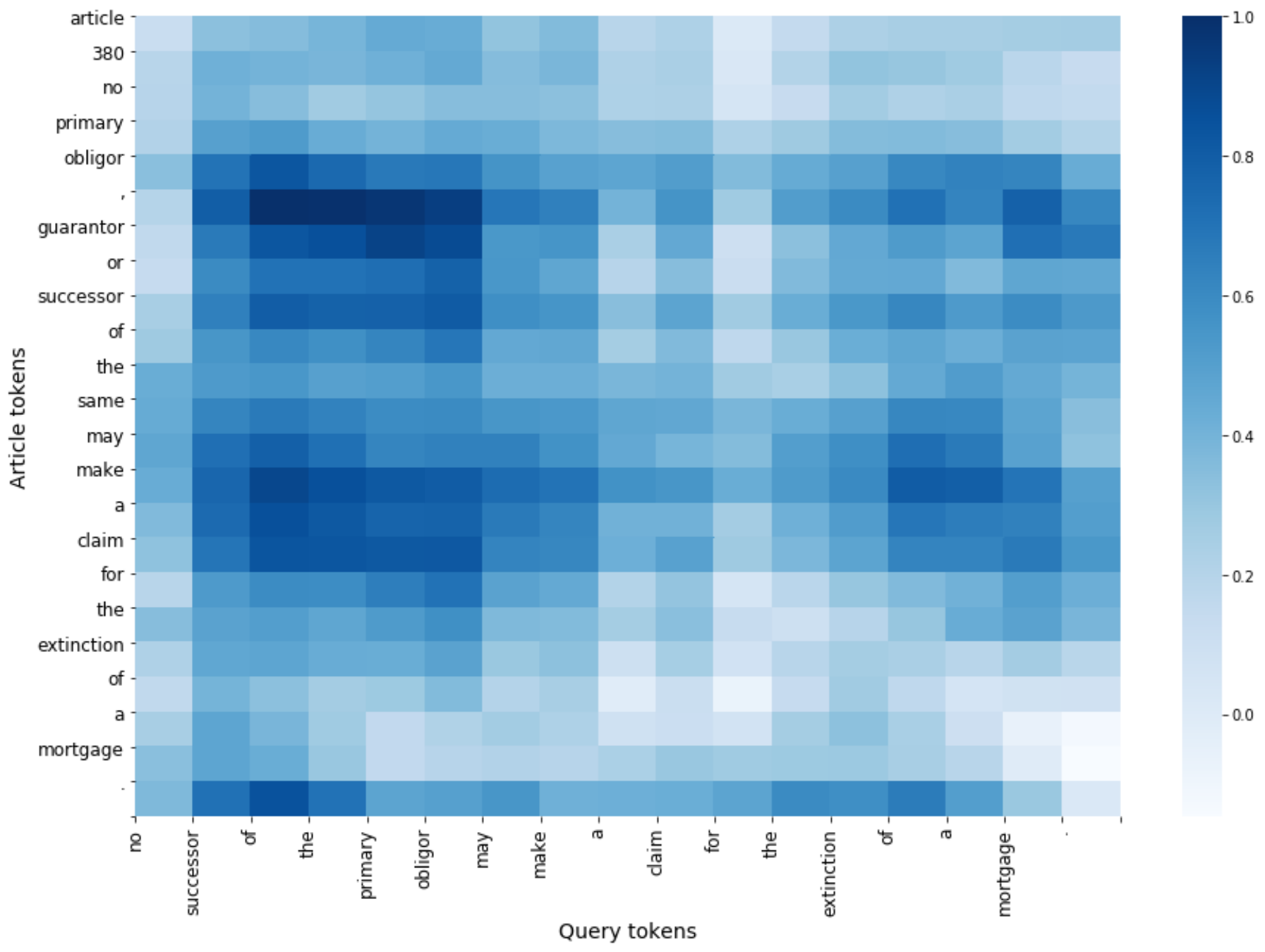}
	\caption{Neural attention example for task H30-13-I. Our model correctly predicted a positive entailment relationship.}
	\Description{Neural attention example for task H30-13-I.}
	\label{fig:att13i}
\end{figure*}
The assessment from the COLIEE competition shown in Table \ref{tab:ent} resulted in an accuracy of our model of 57.14\%, compared to the best result of 68.37\% and the lowest score of 44.9\%. The baseline which always predicts a negative entailment achieved 52.04\% accuracy. Since our model works with neural attention, we can visualize the query-document interaction in our model for selected cases. For instance, in task H30-23-A, our model predicted the positive class, although the correct answer is a negative entailment.

Figure \ref{fig:att23a} illustrates the neuron activation of one of our stacked encoder models. Note that there is almost no neuron activation between the query term ``pre-contract'' and the article tokens. We suppose that the Glove embeddings do not provide feature information for the legal use of this specific hyphenated term. Interestingly, there is a high neuron activation between the query sequence ``shall be effective'' and  ``when one of the parties receives'', as well as the article number within the law document. The sequence ``that are the same in kind'' also exhibits higher activation values for that query sequence. Therefore, we attribute this mistake to the out-of-vocabulary-problem, which is often encountered in domain-specific texts with scarce resources for context modeling with embeddings. While own Word2Vec embeddings were employed for the information retrieval task, we selected Glove embeddings because of the presumably higher likelihood of correctly learned semantic feature representations for negation words and terms such as ``must'' and ``may''. 

The second example in Figure \ref{fig:att13i} shows a correctly classified positive entailment relationship. Elevated activation scores are found in the query terms ``no successor of the primary obligor may'' in conjunction with the article term sequence ``no primary obligor, guarantor or successor'' and ``may make a claim for the extinction''. In this case, it can also be observed that the article number has not been attended as strong as in the previous example.    

The previous two examples show that our stacked encoder has the ability to attend meaningful inputs, given that the input feature representation is expressive enough to capture the domain-specific context of a word. Due to space limitations, examples with positive predictions were chosen because both encoders are highly biased towards the negative class. In the few cases where they predict a positive entailment, they appear to have a local competence for correctly identifying vital features. Therefore, we expect the performance to increase when word embeddings are generated from a large domain-specific corpus. 
\section{Conclusion}
In this work, we present a law retrieval approach and an entailment classifier, developed during the COLIEE '19 competition. For the law retrieval task, we combined BM25 scoring with the Word Centroid Distance for word embeddings, and then applied similarity thresholding for a variable number of retrieved documents per query. Our results outperform some competitor approaches, but we expect further improvements by using word embeddings trained on a larger legal corpus. Regarding the textual entailment task, we outperformed the baseline and thereby have shown a benefit of using a stacked encoder architecture with additional features and an attention layer. Future work for the entailment task can be done on additional feature extraction, for instance semantic role labels. Since our ensemble has been selected manually, it can be beneficial to use automated pruning approaches which are designed for imbalanced data to overcome classifier bias and exploit complementary local competences \cite{krawczyk2018leveraging}. Another research field is the use of different learning architectures and other word embeddings suitable for the legal domain.    

\bibliographystyle{ACM-Reference-Format}
\bibliography{COLIEE19}


\begin{thebibliography}{34}


\ifx \showCODEN    \undefined \def \showCODEN     #1{\unskip}     \fi
\ifx \showDOI      \undefined \def \showDOI       #1{#1}\fi
\ifx \showISBNx    \undefined \def \showISBNx     #1{\unskip}     \fi
\ifx \showISBNxiii \undefined \def \showISBNxiii  #1{\unskip}     \fi
\ifx \showISSN     \undefined \def \showISSN      #1{\unskip}     \fi
\ifx \showLCCN     \undefined \def \showLCCN      #1{\unskip}     \fi
\ifx \shownote     \undefined \def \shownote      #1{#1}          \fi
\ifx \showarticletitle \undefined \def \showarticletitle #1{#1}   \fi
\ifx \showURL      \undefined \def \showURL       {\relax}        \fi
\providecommand\bibfield[2]{#2}
\providecommand\bibinfo[2]{#2}
\providecommand\natexlab[1]{#1}
\providecommand\showeprint[2][]{arXiv:#2}

\bibitem[\protect\citeauthoryear{Arora, Hossari, Maldonado, Conran, Paulus, and
  Dirschl}{Arora et~al\mbox{.}}{2018}]%
        {arora2018}
\bibfield{author}{\bibinfo{person}{Piyush Arora}, \bibinfo{person}{Murhaf
  Hossari}, \bibinfo{person}{Alfredo Maldonado}, \bibinfo{person}{Gareth~J.F.
  Conran, Clare~andJones}, \bibinfo{person}{Johannes Paulus,
  Alexander~andKlostermann}, {and} \bibinfo{person}{Christian Dirschl}.}
  \bibinfo{year}{2018}\natexlab{}.
\newblock \showarticletitle{Challenges in the Development of Effective Systems
  for Professional Legal Search}. In
  \bibinfo{booktitle}{\emph{ProfS/KG4IR/Data: Search@ SIGIR}}.
  \bibinfo{publisher}{CEUR-WS.org}, \bibinfo{address}{Ann Arbor, Michigan,
  USA}, \bibinfo{pages}{29--34}.
\newblock


\bibitem[\protect\citeauthoryear{Bahdanau, Cho, and Bengio}{Bahdanau
  et~al\mbox{.}}{2014}]%
        {Bahdanau2015NeuralMT}
\bibfield{author}{\bibinfo{person}{Dzmitry Bahdanau},
  \bibinfo{person}{Kyunghyun Cho}, {and} \bibinfo{person}{Yoshua Bengio}.}
  \bibinfo{year}{2014}\natexlab{}.
\newblock \showarticletitle{Neural machine translation by jointly learning to
  align and translate}.
\newblock \bibinfo{journal}{\emph{arXiv preprint arXiv:1409.0473}}
  (\bibinfo{year}{2014}).
\newblock


\bibitem[\protect\citeauthoryear{Bowman, Angeli, Potts, and Manning}{Bowman
  et~al\mbox{.}}{2015}]%
        {bowman2015large}
\bibfield{author}{\bibinfo{person}{Samuel~R. Bowman}, \bibinfo{person}{Gabor
  Angeli}, \bibinfo{person}{Christopher Potts}, {and}
  \bibinfo{person}{Christopher~D. Manning}.} \bibinfo{year}{2015}\natexlab{}.
\newblock \showarticletitle{A large annotated corpus for learning natural
  language inference}. In \bibinfo{booktitle}{\emph{Proceedings of the 2015
  Conference on Empirical Methods in Natural Language Processing}}.
  \bibinfo{publisher}{Association for Computational Linguistics},
  \bibinfo{address}{Lisbon, Portugal}, \bibinfo{pages}{632--642}.
\newblock
\urldef\tempurl%
\url{https://doi.org/10.18653/v1/D15-1075}
\showDOI{\tempurl}


\bibitem[\protect\citeauthoryear{Carvalho, Nguyen, Tran, and Nguyen}{Carvalho
  et~al\mbox{.}}{2017}]%
        {carvalho2015lexical}
\bibfield{author}{\bibinfo{person}{Danilo~S. Carvalho},
  \bibinfo{person}{Minh-Tien Nguyen}, \bibinfo{person}{Chien-Xuan Tran}, {and}
  \bibinfo{person}{Minh-Le Nguyen}.} \bibinfo{year}{2017}\natexlab{}.
\newblock \showarticletitle{Lexical-Morphological Modeling for Legal Text
  Analysis}. In \bibinfo{booktitle}{\emph{New Frontiers in Artificial
  Intelligence}}, \bibfield{editor}{\bibinfo{person}{Mihoko Otake},
  \bibinfo{person}{Setsuya Kurahashi}, \bibinfo{person}{Yuiko Ota},
  \bibinfo{person}{Ken Satoh}, {and} \bibinfo{person}{Daisuke Bekki}} (Eds.).
  \bibinfo{publisher}{Springer International Publishing},
  \bibinfo{address}{Cham}, \bibinfo{pages}{295--311}.
\newblock
\showISBNx{978-3-319-50953-2}


\bibitem[\protect\citeauthoryear{Chen, Fisch, Weston, and Bordes}{Chen
  et~al\mbox{.}}{2017}]%
        {chen2017reading}
\bibfield{author}{\bibinfo{person}{Danqi Chen}, \bibinfo{person}{Adam Fisch},
  \bibinfo{person}{Jason Weston}, {and} \bibinfo{person}{Antoine Bordes}.}
  \bibinfo{year}{2017}\natexlab{}.
\newblock \showarticletitle{Reading {W}ikipedia to Answer Open-Domain
  Questions}. In \bibinfo{booktitle}{\emph{Proceedings of the 55th Annual
  Meeting of the Association for Computational Linguistics (Volume 1: Long
  Papers)}}. \bibinfo{publisher}{Association for Computational Linguistics},
  \bibinfo{address}{Vancouver, Canada}, \bibinfo{pages}{1870--1879}.
\newblock
\urldef\tempurl%
\url{https://doi.org/10.18653/v1/P17-1171}
\showDOI{\tempurl}


\bibitem[\protect\citeauthoryear{Do, Nguyen, Tran, Nguyen, and Nguyen}{Do
  et~al\mbox{.}}{2017}]%
        {DoNTNN17}
\bibfield{author}{\bibinfo{person}{Phong-Khac Do}, \bibinfo{person}{Huy-Tien
  Nguyen}, \bibinfo{person}{Chien-Xuan Tran}, \bibinfo{person}{Minh-Tien
  Nguyen}, {and} \bibinfo{person}{Minh-Le Nguyen}.}
  \bibinfo{year}{2017}\natexlab{}.
\newblock \showarticletitle{Legal question answering using ranking SVM and deep
  convolutional neural network}.
\newblock \bibinfo{journal}{\emph{arXiv preprint arXiv:1703.05320}}
  (\bibinfo{year}{2017}).
\newblock


\bibitem[\protect\citeauthoryear{Galke, Saleh, and Scherp}{Galke
  et~al\mbox{.}}{2017}]%
        {galke2017word}
\bibfield{author}{\bibinfo{person}{Lukas Galke}, \bibinfo{person}{Ahmed Saleh},
  {and} \bibinfo{person}{Ansgar Scherp}.} \bibinfo{year}{2017}\natexlab{}.
\newblock \showarticletitle{Word Embeddings for Practical Information
  Retrieval}.
\newblock \bibinfo{journal}{\emph{INFORMATIK 2017}} (\bibinfo{year}{2017}),
  \bibinfo{pages}{2155--2167}.
\newblock


\bibitem[\protect\citeauthoryear{Glickman and Dagan}{Glickman and
  Dagan}{2005}]%
        {glickman2005web}
\bibfield{author}{\bibinfo{person}{Oren Glickman} {and} \bibinfo{person}{Ido
  Dagan}.} \bibinfo{year}{2005}\natexlab{}.
\newblock \showarticletitle{Web based probabilistic textual entailment}. In
  \bibinfo{booktitle}{\emph{In Proceedings of the 1st Pascal Challenge
  Workshop}}. \bibinfo{pages}{33--36}.
\newblock


\bibitem[\protect\citeauthoryear{Ha, Ha, Nguyen, and Thi}{Ha
  et~al\mbox{.}}{2012}]%
        {ha2012refining}
\bibfield{author}{\bibinfo{person}{Quang-Thuy Ha}, \bibinfo{person}{Thi-Oanh
  Ha}, \bibinfo{person}{Thi-Dung Nguyen}, {and}
  \bibinfo{person}{Thuy-Linh~Nguyen Thi}.} \bibinfo{year}{2012}\natexlab{}.
\newblock \showarticletitle{Refining the Judgment Threshold to Improve
  Recognizing Textual Entailment Using Similarity}. In
  \bibinfo{booktitle}{\emph{Computational Collective Intelligence. Technologies
  and Applications}}, \bibfield{editor}{\bibinfo{person}{Ngoc-Thanh Nguyen},
  \bibinfo{person}{Kiem Hoang}, {and} \bibinfo{person}{Piotr Jedrzejowicz}}
  (Eds.). \bibinfo{publisher}{Springer Berlin Heidelberg},
  \bibinfo{address}{Berlin, Heidelberg}, \bibinfo{pages}{335--344}.
\newblock
\showISBNx{978-3-642-34707-8}


\bibitem[\protect\citeauthoryear{Islam and Inkpen}{Islam and Inkpen}{2008}]%
        {islam2008semantic}
\bibfield{author}{\bibinfo{person}{Aminul Islam} {and} \bibinfo{person}{Diana
  Inkpen}.} \bibinfo{year}{2008}\natexlab{}.
\newblock \showarticletitle{Semantic text similarity using corpus-based word
  similarity and string similarity}.
\newblock \bibinfo{journal}{\emph{ACM Transactions on Knowledge Discovery from
  Data (TKDD)}} \bibinfo{volume}{2}, \bibinfo{number}{2}
  (\bibinfo{year}{2008}), \bibinfo{pages}{10}.
\newblock


\bibitem[\protect\citeauthoryear{Kim, Goebel, Kano, and Satoh}{Kim
  et~al\mbox{.}}{2016}]%
        {kim2016coliee}
\bibfield{author}{\bibinfo{person}{Mi-Young Kim}, \bibinfo{person}{Randy
  Goebel}, \bibinfo{person}{Yoshinobu Kano}, {and} \bibinfo{person}{Ken
  Satoh}.} \bibinfo{year}{2016}\natexlab{}.
\newblock \showarticletitle{COLIEE-2016: evaluation of the competition on legal
  information extraction and entailment}. In
  \bibinfo{booktitle}{\emph{International Workshop on Juris-informatics
  (JURISIN 2016)}}.
\newblock


\bibitem[\protect\citeauthoryear{Kim, Goebel, and Satoh}{Kim
  et~al\mbox{.}}{2015}]%
        {kim2015coliee}
\bibfield{author}{\bibinfo{person}{Mi-Young Kim}, \bibinfo{person}{Randy
  Goebel}, {and} \bibinfo{person}{Ken Satoh}.} \bibinfo{year}{2015}\natexlab{}.
\newblock \showarticletitle{COLIEE-2015: evaluation of legal question
  answering}. In \bibinfo{booktitle}{\emph{Ninth International Workshop on
  Juris-informatics (JURISIN 2015)}}.
\newblock


\bibitem[\protect\citeauthoryear{Kim, Lu, Rabelo, and Goebel}{Kim
  et~al\mbox{.}}{2018}]%
        {kimcoliee}
\bibfield{author}{\bibinfo{person}{Mi-Young Kim}, \bibinfo{person}{Yao Lu},
  \bibinfo{person}{Juliano Rabelo}, {and} \bibinfo{person}{Randy Goebel}.}
  \bibinfo{year}{2018}\natexlab{}.
\newblock \bibinfo{title}{COLIEE-2018: Evaluation of the Competition on Case
  Law Information Extraction and Entailment}.
\newblock \bibinfo{howpublished}{[online]}.
\newblock
\urldef\tempurl%
\url{https://sites.ualberta.ca/~rabelo/COLIEE2019/COLIEE2018_CL_summary.pdf}
\showURL{%
\tempurl}


\bibitem[\protect\citeauthoryear{Krawczyk and Wo{\'z}niak}{Krawczyk and
  Wo{\'z}niak}{2018}]%
        {krawczyk2018leveraging}
\bibfield{author}{\bibinfo{person}{Bartosz Krawczyk} {and}
  \bibinfo{person}{Micha{\l} Wo{\'z}niak}.} \bibinfo{year}{2018}\natexlab{}.
\newblock \showarticletitle{Leveraging Ensemble Pruning for Imbalanced Data
  Classification}. In \bibinfo{booktitle}{\emph{2018 IEEE International
  Conference on Systems, Man, and Cybernetics (SMC)}}.
  \bibinfo{publisher}{{IEEE}}, \bibinfo{pages}{439--444}.
\newblock


\bibitem[\protect\citeauthoryear{Kusner, Sun, Kolkin, and Weinberger}{Kusner
  et~al\mbox{.}}{2015}]%
        {kusner2015word}
\bibfield{author}{\bibinfo{person}{Matt~J. Kusner}, \bibinfo{person}{Yu Sun},
  \bibinfo{person}{Nicholas~I. Kolkin}, {and} \bibinfo{person}{Kilian~Q.
  Weinberger}.} \bibinfo{year}{2015}\natexlab{}.
\newblock \showarticletitle{From Word Embeddings to Document Distances}. In
  \bibinfo{booktitle}{\emph{Proceedings of the 32Nd International Conference on
  International Conference on Machine Learning - Volume 37}}
  \emph{(\bibinfo{series}{ICML'15})}. \bibinfo{publisher}{JMLR.org},
  \bibinfo{pages}{957--966}.
\newblock
\urldef\tempurl%
\url{http://dl.acm.org/citation.cfm?id=3045118.3045221}
\showURL{%
\tempurl}


\bibitem[\protect\citeauthoryear{Liu, Sun, Lin, and Wang}{Liu
  et~al\mbox{.}}{2016}]%
        {Liu2016LearningNL}
\bibfield{author}{\bibinfo{person}{Yang Liu}, \bibinfo{person}{Chengjie Sun},
  \bibinfo{person}{Lei Lin}, {and} \bibinfo{person}{Xiaolong Wang}.}
  \bibinfo{year}{2016}\natexlab{}.
\newblock \showarticletitle{Learning natural language inference using
  bidirectional LSTM model and inner-attention}.
\newblock \bibinfo{journal}{\emph{arXiv preprint arXiv:1605.09090}}
  (\bibinfo{year}{2016}).
\newblock


\bibitem[\protect\citeauthoryear{Mandal, Ghosh, Bhattacharya, Pal, and
  Ghosh}{Mandal et~al\mbox{.}}{2017}]%
        {mandal2017overview}
\bibfield{author}{\bibinfo{person}{Arpan Mandal}, \bibinfo{person}{Kripabandhu
  Ghosh}, \bibinfo{person}{Arnab Bhattacharya}, \bibinfo{person}{Arindam Pal},
  {and} \bibinfo{person}{Saptarshi Ghosh}.} \bibinfo{year}{2017}\natexlab{}.
\newblock \showarticletitle{Overview of the FIRE 2017 IRLeD Track: Information
  Retrieval from Legal Documents}. In \bibinfo{booktitle}{\emph{FIRE (Working
  Notes)}}, \bibfield{editor}{\bibinfo{person}{Prasenjit Majumder},
  \bibinfo{person}{Mandar Mitra}, \bibinfo{person}{Parth~Mehta Mehta}, {and}
  \bibinfo{person}{Jainisha Sankhavara}} (Eds.).
  \bibinfo{publisher}{CEUR-WS.org}, \bibinfo{address}{Bangalore, India},
  \bibinfo{pages}{63--68}.
\newblock


\bibitem[\protect\citeauthoryear{Manning, Raghavan, and Sch\"{u}tze}{Manning
  et~al\mbox{.}}{2008}]%
        {manning2010introduction}
\bibfield{author}{\bibinfo{person}{Christopher~D. Manning},
  \bibinfo{person}{Prabhakar Raghavan}, {and} \bibinfo{person}{Hinrich
  Sch\"{u}tze}.} \bibinfo{year}{2008}\natexlab{}.
\newblock \bibinfo{booktitle}{\emph{Introduction to Information Retrieval}}.
\newblock \bibinfo{publisher}{Cambridge University Press},
  \bibinfo{address}{New York, NY, USA}.
\newblock
\showISBNx{0521865719, 9780521865715}


\bibitem[\protect\citeauthoryear{Mikolov, Sutskever, Chen, Corrado, and
  Dean}{Mikolov et~al\mbox{.}}{2013}]%
        {mikolov2013distributed}
\bibfield{author}{\bibinfo{person}{Tomas Mikolov}, \bibinfo{person}{Ilya
  Sutskever}, \bibinfo{person}{Kai Chen}, \bibinfo{person}{Greg Corrado}, {and}
  \bibinfo{person}{Jeffrey Dean}.} \bibinfo{year}{2013}\natexlab{}.
\newblock \showarticletitle{Distributed Representations of Words and Phrases
  and Their Compositionality}. In \bibinfo{booktitle}{\emph{Proceedings of the
  26th International Conference on Neural Information Processing Systems -
  Volume 2}} \emph{(\bibinfo{series}{NIPS'13})}. \bibinfo{publisher}{Curran
  Associates Inc.}, \bibinfo{address}{USA}, \bibinfo{pages}{3111--3119}.
\newblock
\urldef\tempurl%
\url{http://dl.acm.org/citation.cfm?id=2999792.2999959}
\showURL{%
\tempurl}


\bibitem[\protect\citeauthoryear{Nanda, John, Caro, Boella, and Robaldo}{Nanda
  et~al\mbox{.}}{2017}]%
        {nanda2017legal}
\bibfield{author}{\bibinfo{person}{Rohan Nanda},
  \bibinfo{person}{Adebayo~Kolawole John}, \bibinfo{person}{Luigi~Di Caro},
  \bibinfo{person}{Guido Boella}, {and} \bibinfo{person}{Livio Robaldo}.}
  \bibinfo{year}{2017}\natexlab{}.
\newblock \showarticletitle{Legal Information Retrieval Using Topic Clustering
  and Neural Networks}. In \bibinfo{booktitle}{\emph{COLIEE 2017. 4th
  Competition on Legal Information Extraction and Entailment}}
  \emph{(\bibinfo{series}{EPiC Series in Computing})},
  \bibfield{editor}{\bibinfo{person}{Ken Satoh}, \bibinfo{person}{Mi-Young
  Kim}, \bibinfo{person}{Yoshinobu Kano}, \bibinfo{person}{Randy Goebel}, {and}
  \bibinfo{person}{Tiago Oliveira}} (Eds.), Vol.~\bibinfo{volume}{47}.
  \bibinfo{publisher}{EasyChair}, \bibinfo{pages}{68--78}.
\newblock
\showISSN{2398-7340}
\urldef\tempurl%
\url{https://doi.org/10.29007/psgx}
\showDOI{\tempurl}


\bibitem[\protect\citeauthoryear{Nie and Bansal}{Nie and Bansal}{2017}]%
        {nie2017shortcut}
\bibfield{author}{\bibinfo{person}{Yixin Nie} {and} \bibinfo{person}{Mohit
  Bansal}.} \bibinfo{year}{2017}\natexlab{}.
\newblock \showarticletitle{Shortcut-Stacked Sentence Encoders for Multi-Domain
  Inference}. In \bibinfo{booktitle}{\emph{Proceedings of the 2nd Workshop on
  Evaluating Vector Space Representations for {NLP}}}.
  \bibinfo{publisher}{Association for Computational Linguistics},
  \bibinfo{address}{Copenhagen, Denmark}, \bibinfo{pages}{41--45}.
\newblock
\urldef\tempurl%
\url{https://doi.org/10.18653/v1/W17-5308}
\showDOI{\tempurl}


\bibitem[\protect\citeauthoryear{Pennington, Socher, and Manning}{Pennington
  et~al\mbox{.}}{2014}]%
        {pennington2014glove}
\bibfield{author}{\bibinfo{person}{Jeffrey Pennington},
  \bibinfo{person}{Richard Socher}, {and} \bibinfo{person}{Christopher
  Manning}.} \bibinfo{year}{2014}\natexlab{}.
\newblock \showarticletitle{{G}love: Global Vectors for Word Representation}.
  In \bibinfo{booktitle}{\emph{Proceedings of the 2014 Conference on Empirical
  Methods in Natural Language Processing ({EMNLP})}}.
  \bibinfo{publisher}{Association for Computational Linguistics},
  \bibinfo{address}{Doha, Qatar}, \bibinfo{pages}{1532--1543}.
\newblock
\urldef\tempurl%
\url{https://doi.org/10.3115/v1/D14-1162}
\showDOI{\tempurl}


\bibitem[\protect\citeauthoryear{Rajpurkar, Zhang, Lopyrev, and
  Liang}{Rajpurkar et~al\mbox{.}}{2016}]%
        {rajpurkar2016squad}
\bibfield{author}{\bibinfo{person}{Pranav Rajpurkar}, \bibinfo{person}{Jian
  Zhang}, \bibinfo{person}{Konstantin Lopyrev}, {and} \bibinfo{person}{Percy
  Liang}.} \bibinfo{year}{2016}\natexlab{}.
\newblock \showarticletitle{{SQ}u{AD}: 100,000+ Questions for Machine
  Comprehension of Text}. In \bibinfo{booktitle}{\emph{Proceedings of the 2016
  Conference on Empirical Methods in Natural Language Processing}}.
  \bibinfo{publisher}{Association for Computational Linguistics},
  \bibinfo{address}{Austin, Texas}, \bibinfo{pages}{2383--2392}.
\newblock
\urldef\tempurl%
\url{https://doi.org/10.18653/v1/D16-1264}
\showDOI{\tempurl}


\bibitem[\protect\citeauthoryear{Reddy, Chen, and Manning}{Reddy
  et~al\mbox{.}}{2018}]%
        {reddy2018coqa}
\bibfield{author}{\bibinfo{person}{Siva Reddy}, \bibinfo{person}{Danqi Chen},
  {and} \bibinfo{person}{Christopher~D Manning}.}
  \bibinfo{year}{2018}\natexlab{}.
\newblock \showarticletitle{Coqa: A conversational question answering
  challenge}.
\newblock \bibinfo{journal}{\emph{arXiv preprint arXiv:1808.07042}}
  (\bibinfo{year}{2018}).
\newblock


\bibitem[\protect\citeauthoryear{Robertson and Zaragoza}{Robertson and
  Zaragoza}{2009}]%
        {robertson2009probabilistic}
\bibfield{author}{\bibinfo{person}{Stephen Robertson} {and}
  \bibinfo{person}{Hugo Zaragoza}.} \bibinfo{year}{2009}\natexlab{}.
\newblock \showarticletitle{The Probabilistic Relevance Framework: BM25 and
  Beyond}.
\newblock \bibinfo{journal}{\emph{Information Retrieval}} \bibinfo{volume}{3},
  \bibinfo{number}{4} (\bibinfo{year}{2009}), \bibinfo{pages}{333--389}.
\newblock


\bibitem[\protect\citeauthoryear{Rockt{\"a}schel, Grefenstette, Hermann,
  Ko{\v{c}}isk{\`y}, and Blunsom}{Rockt{\"a}schel et~al\mbox{.}}{2015}]%
        {Rocktschel2016ReasoningAE}
\bibfield{author}{\bibinfo{person}{Tim Rockt{\"a}schel},
  \bibinfo{person}{Edward Grefenstette}, \bibinfo{person}{Karl~Moritz Hermann},
  \bibinfo{person}{Tom{\'a}{\v{s}} Ko{\v{c}}isk{\`y}}, {and}
  \bibinfo{person}{Phil Blunsom}.} \bibinfo{year}{2015}\natexlab{}.
\newblock \showarticletitle{Reasoning about entailment with neural attention}.
\newblock \bibinfo{journal}{\emph{arXiv preprint arXiv:1509.06664}}
  (\bibinfo{year}{2015}).
\newblock


\bibitem[\protect\citeauthoryear{Rooney, Wang, and Taylor}{Rooney
  et~al\mbox{.}}{2014}]%
        {rooney2014investigation}
\bibfield{author}{\bibinfo{person}{Niall Rooney}, \bibinfo{person}{Hui Wang},
  {and} \bibinfo{person}{Philip~S Taylor}.} \bibinfo{year}{2014}\natexlab{}.
\newblock \showarticletitle{An investigation into the application of ensemble
  learning for entailment classification}.
\newblock \bibinfo{journal}{\emph{Information Processing \& Management}}
  \bibinfo{volume}{50}, \bibinfo{number}{1} (\bibinfo{year}{2014}),
  \bibinfo{pages}{87--103}.
\newblock


\bibitem[\protect\citeauthoryear{Schmidt-Hieber}{Schmidt-Hieber}{2017}]%
        {schmidt2017nonparametric}
\bibfield{author}{\bibinfo{person}{Johannes Schmidt-Hieber}.}
  \bibinfo{year}{2017}\natexlab{}.
\newblock \showarticletitle{Nonparametric regression using deep neural networks
  with ReLU activation function}.
\newblock \bibinfo{journal}{\emph{arXiv preprint arXiv:1708.06633}}
  (\bibinfo{year}{2017}).
\newblock


\bibitem[\protect\citeauthoryear{Stein, zu~Eissen, and Potthast}{Stein
  et~al\mbox{.}}{2007}]%
        {stein2007strategies}
\bibfield{author}{\bibinfo{person}{Benno Stein}, \bibinfo{person}{Sven~Meyer zu
  Eissen}, {and} \bibinfo{person}{Martin Potthast}.}
  \bibinfo{year}{2007}\natexlab{}.
\newblock \showarticletitle{Strategies for Retrieving Plagiarized Documents}.
  In \bibinfo{booktitle}{\emph{Proceedings of the 30th Annual International ACM
  SIGIR Conference on Research and Development in Information Retrieval}}
  \emph{(\bibinfo{series}{SIGIR '07})}. \bibinfo{publisher}{ACM},
  \bibinfo{address}{New York, NY, USA}, \bibinfo{pages}{825--826}.
\newblock
\showISBNx{978-1-59593-597-7}
\urldef\tempurl%
\url{https://doi.org/10.1145/1277741.1277928}
\showDOI{\tempurl}


\bibitem[\protect\citeauthoryear{Stroh and Mathur}{Stroh and Mathur}{2016}]%
        {strohquestion}
\bibfield{author}{\bibinfo{person}{Eylon Stroh} {and} \bibinfo{person}{Priyank
  Mathur}.} \bibinfo{year}{2016}\natexlab{}.
\newblock \bibinfo{title}{Question Answering Using Deep Learning}.
\newblock \bibinfo{howpublished}{[online]}.
\newblock
\urldef\tempurl%
\url{https://cs224d.stanford.edu/reports/StrohMathur.pdf}
\showURL{%
\tempurl}


\bibitem[\protect\citeauthoryear{Sutskever, Vinyals, and Le}{Sutskever
  et~al\mbox{.}}{2014}]%
        {sutskever2014sequence}
\bibfield{author}{\bibinfo{person}{Ilya Sutskever}, \bibinfo{person}{Oriol
  Vinyals}, {and} \bibinfo{person}{Quoc~V. Le}.}
  \bibinfo{year}{2014}\natexlab{}.
\newblock \showarticletitle{Sequence to Sequence Learning with Neural
  Networks}. In \bibinfo{booktitle}{\emph{Proceedings of the 27th International
  Conference on Neural Information Processing Systems - Volume 2}}
  \emph{(\bibinfo{series}{NIPS'14})}. \bibinfo{publisher}{MIT Press},
  \bibinfo{address}{Cambridge, MA, USA}, \bibinfo{pages}{3104--3112}.
\newblock
\urldef\tempurl%
\url{http://dl.acm.org/citation.cfm?id=2969033.2969173}
\showURL{%
\tempurl}


\bibitem[\protect\citeauthoryear{Tian, Ning, Kong, Han, Xiao, and Qi}{Tian
  et~al\mbox{.}}{2017}]%
        {tian2017hljit2017}
\bibfield{author}{\bibinfo{person}{Liuyang Tian}, \bibinfo{person}{Hui Ning},
  \bibinfo{person}{Leilei Kong}, \bibinfo{person}{Zhongyuan Han},
  \bibinfo{person}{Ruiming Xiao}, {and} \bibinfo{person}{Haoliang Qi}.}
  \bibinfo{year}{2017}\natexlab{}.
\newblock \showarticletitle{HLJIT2017@ IRLed-FIRE2017: Information Retrieval
  From Legal Documents}. In \bibinfo{booktitle}{\emph{FIRE (Working Notes)}},
  \bibfield{editor}{\bibinfo{person}{Prasenjit Majumder},
  \bibinfo{person}{Mandar Mitra}, \bibinfo{person}{Parth~Mehta Mehta}, {and}
  \bibinfo{person}{Jainisha Sankhavara}} (Eds.).
  \bibinfo{publisher}{CEUR-WS.org}, \bibinfo{address}{Bangalore, India},
  \bibinfo{pages}{82--85}.
\newblock


\bibitem[\protect\citeauthoryear{Yarotsky}{Yarotsky}{2018}]%
        {yarotsky2018optimal}
\bibfield{author}{\bibinfo{person}{Dmitry Yarotsky}.}
  \bibinfo{year}{2018}\natexlab{}.
\newblock \showarticletitle{Optimal approximation of continuous functions by
  very deep ReLU networks}.
\newblock \bibinfo{journal}{\emph{arXiv preprint arXiv:1802.03620}}
  (\bibinfo{year}{2018}).
\newblock


\bibitem[\protect\citeauthoryear{Yoshioka, Kano, Kiyota, and Satoh}{Yoshioka
  et~al\mbox{.}}{2018}]%
        {yoshiokaoverview}
\bibfield{author}{\bibinfo{person}{Masaharu Yoshioka},
  \bibinfo{person}{Yoshinobu Kano}, \bibinfo{person}{Naoki Kiyota}, {and}
  \bibinfo{person}{Ken Satoh}.} \bibinfo{year}{2018}\natexlab{}.
\newblock \bibinfo{title}{Overview of Japanese Statute Law Retrieval and
  Entailment Task at COLIEE-2018}.
\newblock \bibinfo{howpublished}{[online]}.
\newblock
\urldef\tempurl%
\url{https://sites.ualberta.ca/~rabelo/COLIEE2019/COLIEE2018_SL_summary.pdf}
\showURL{%
\tempurl}


\end{thebibliography}
\end{document}